%Paper: 9205002
%From: f77497a <f77497a@kyu-cc.cc.kyushu-u.ac.jp>
%Date: Fri, 1 May 92 19:15:36 JST

%%%%%%%%%%%%%%%%%%%%%%%%%%%%%%%%%%%%%%%
% This uses macros harvmac and epsf.
%%%%%%%%%%%%%%%%%%%%%%%%%%%%%%%%%%%%%%%
\input harvmac
\input epsf
%%%%%%%%%%%%%%%%%%%%%%%%%%%%%%%%% journals %%%%%%%%%%%%%%%%%%%%%%%%%%%%%%%%
\def\AP#1{Ann.\ Phys. {\bf{#1}}}

\def\NP#1{Nucl.\ Phys. {\bf B{#1}}}

\def\PRL#1{Phys.\ Rev.\ Lett. {\bf {#1}}}
\def\PTP#1{Prog.\ Theor.\ Phys. {\bf {#1}}}

%%%%%%%%%%%%%%%%%%%%%%%%%%%%%%%%%%%%%%%%%%%%%%%%%%%%%%%%%%%%%%%%%%%%%
\def\gtnoteq{\mathrel{\hbox{\raise0.2ex
    \hbox{$>$}\kern-0.75em\raise-0.9ex\hbox{$\not=$}}}} \def\Ar{{\cal A}}
\def\De#1{D_\epsilon( #1 )} \def\Dr#1{D_r( #1 )} %\def\bpsi{\bar\psi}
\def\Aa{{\cal A}}
\def\bpsi{\bar\psi}
%%%%%--------------------------------------------------
% The following is for Mac
%\def\graph{\lower2cm
%\vbox to 4.5cm{\hrule width 4.8cm height 0pt depth 0pt
%      \vfill\special{illustration pic.ps}}}
%%%%%--------------------------------------------------

%%%%  title page   %%%%%%

\Title{\vbox{\baselineskip12pt\hbox{KYUSHU--HET--4}\hbox{SAGA--HE--42}}}
{{\vbox{\centerline{ Chiral Invariance and Species Doublers}\vskip10pt
               \centerline{in Generic Fermion Models on the Lattice} }}}
\centerline{Koichi Funakubo$^*$ and Taro Kashiwa} \bigskip
\centerline{${}^*$Department of Physics, Saga University, Saga, 840 JAPAN}
\smallskip \centerline{Department of Physics, Kyushu University, Fukuoka,
812 JAPAN}

\vskip1cm

Discussions are made on the structures of chirally invariant
lattice actions without any restriction of hermiticity.
With the help of the
Ward-Takahashi identity a general conclusion can
be derived that there must
be species doublers in any chirally invariant model
provided that the model is
chosen as well-regularized, that is, there is no
singularity in the propagator
after introducing fermion mass on the lattice.
Various examples are discussed to
pick up better models defined in the sense that the number of species
doubler is smaller than that of the naive Dirac action.

%\draft
\Date{4/92}
\newsec{Introduction}

Since the advent of the no-go theorem of Nielsen and
Ninomiya \ref\NN{H. B. Nielsen and M. Ninomiya, Nucl. Phys. {\bf B185}, 20
(1981); {\bf B193}, 173 (1981); (E) {\bf B193} , 541 (1981).},
people has struggled
to put handed fermions on the
lattice\ref\GM{M. F. L. Golterman, Nuc. Phys. {\bf
B} (Proc. Suppl.) {\bf 20}, 528 (1991); I. Montvay, Plenary Talk at Lattice
'91(to be published in Nucl. Phys. {\bf B} (Proc. Suppl.) )}.
If this would be done,
the nonperturbative treatment becomes possible to obtain the top quark mass
and the baryon number generation in the standard model and to reduce some
problems in the technicolor models.  The theorem tells us that any chirally
symmetric action with (i) locality (ii) translational invariance and (iii)
hermiticity must always have equal number of left and right handed fermions
(species doublers).
In other words, we cannot help breaking a chiral invariance
if we throw away those unwanted
particles: Wilson \ref\WI{K. G. Wilson, in {\sl
New Phenomena in Subnuclear Physics},  e.d. A. Zichichi, Erice Lecture 1975,
(Plenum 1977).} introduced the so-called Wilson term which breaks the chiral
symmetry.  The situation is the same in the case of Majorana-type
fermion\ref\CP{C. Pryor, Phys. Rev. {\bf D43}, 2669 (1991).}. So far attempts
have been made to lift the conditions (i)
and/or (ii): the introduction of (Higgs)
scalars \ref\SS{P. D. V. Swift, Phys. Lett. {\bf B145}, 256 (1984);
J. Smit, Acta
Physica Polonica {\bf B17}, 531 (1986).}\ref\FK{K. Funakubo and T. Kashiwa,
Phys. Rev. Lett,{\bf 60}, 2113 (1988).}%
\ref\SA{S. Aoki, Phys. Rev. Lett. {\bf 60},
2109 (1988).} to the fermion action, yielding chiral gauge models, can be
regarded as a non-local action after being
integrated out with respect to those
scalars. (Also there exists free (right) handed fermion in these
approachs\ref\GP{M. F. L. Golterman and D. N. Petcher, Phys. Lett. {\bf B225}
159 (1989)}.) The adoption of a random
lattice can be classified in a similar class
since in this approach the random variable behaves as
a scalar field \ref\RM{T.
W. Chiu, Phys. Lett. {\bf B217}, 151 (1989); {\bf B206}, 510 (1988); S. J.
Perantonius and J. F. Wheater, Nucl. Phys.
{\bf B295}, [FS 21], 443 (1988). }.  The
mirror fermion method \ref\MO{I. Montvay, Phys. Lett.
{\bf B199}, 89 (1987).},
uses an additional (mirror) fermion, which leads
us again a non-local action
after the integration of the mirror fermion(again with free redundant
fermions).

Contrary to the above, there have been very few serious efforts \ref\GLR{M.
Gross, G. P. Lepage and P. E. L. Rakow,
Phys. Lett. {\bf B197}, 183 (1987).} for
lifting the condition (iii); hermiticity.
The reason is that it is hard to deal with
functions of complex variables if we abandon the hermiticity. Indeed it is
very difficult to prove the no-go theorem in nonhermitian cases. However it
would be economical in the sense that there
is no need for the introduction of
additional degrees of freedom to get a chirally symmetric model by throwing
away the hermiticity. In this paper, we study the structure of chirally
invariant lattice action with the help of Ward-Takahashi identity\ref\FKa{K.
Funakubo and T. Kashiwa, to be published in Nucl. Phys. {\bf B} (Lattice '91
Proc. Suppl.)}.  So far the existence of
species doublers in a chirally symmetric
and nonhermitian model has been anticipated by ref.\GLR\ but we need a more
general argument. Our strategy is as follows:

\item{} Step 1: Knowing that; propagator behaves $\sum_{\mu} i\gamma_\mu
p_\mu + M $ around $p = 0$ whose contribution to the Ward-Takahashi identity
gives a well-known anomaly \ref\KSKS{ L. H. Karsten and J. Smit,  \NP{183},
103 (1981); T. Kashiwa and H. So, \PTP{73}, 762 (1985).}

\item{} Step 2: Knowing that; on
the lattice any chirally symmetric model does
not have anomaly.

\item{} Step 3: Thus there must be the other zero of the inverse propagator
which cancels the anomaly from $p=0$. (This is species doubler.)

This leads us to the conclusion that if a model is well-regularized there
should be species doublers in any chirally invariant model,
which generalizes
the no-go theorem to include nonhermitian actions.

In section 2, we set up a general form of
chirally invariant actions and list
some examples. In section 3, the Ward-Takahashi identity and the way to the
continuum limit are discussed. Detailed calculations in two dimensions are
then performed for general models with chiral symmetry in section 4
to illustrate
our conclusion. The final  section is
devoted to further discussions. In the
appendix, we present a brief introduction of
the reflection positivity which is
necessary to define the hermiticity on the lattice.

\newsec{General Form of Fermion Actions}

We write a general fermion action in $d$ dimensions as
\eqn\Ia{
\eqalign{
  I= & -\sum_n\sum_{\mu=1}^d \Bigl\{ \sum_{k\geq 1}\bigl[
       \bar\psi(n)\Gamma_\mu^{(+;k)}U_\mu(n)U_\mu(n+\mu)
    \cdots U_\mu(n+(k-1)\mu) \psi(n+k\mu) \cr
     &\quad + \bar\psi(n+k\mu)\Gamma_\mu^{(-;k)}
       U_\mu^\dagger(n+(k-1)\mu)
      U_\mu^\dagger(n+(k-2)\mu)\cdots U_\mu^\dagger(n) \psi(n)
      \bigr] \Bigr\}                \cr
     & - \sum_n \bar\psi(n) \Bigl( \Gamma^{(0)} + M \Bigr)\psi(n),   \cr
}}
where $\Gamma_\mu^{(\pm;k)}$ and $\Gamma^{(0)} $ are made from
$\gamma$-matrices and $k$ is an integer running within some finite range
to satisfy locality. (A more general case may be considered; where
$\bar\psi(n)$ and $\psi(n+\cdots)$ are not located on a straight line
such as $\bar\psi(n)\cdots\psi(n+k_1\mu+k_2\nu)$ $(\mu\not=\nu)$. We
shall not adopt such a model, since the choice of the link variables to
connect them is not unique.) $U_\mu(n)$ is the usual link variable
\eqn\Ib{
   U_\mu(n) = {\rm e}^{iA_\mu(n)},
   }
where the coupling constant has been absorbed in the definition of gauge
fields $A_\mu$. We take all quantities dimensionless such that
\eqn\Ic{
\eqalign{
     \psi(n)  &\equiv a^{(d-1)/2}\tilde\psi(x),        \cr
     A_\mu(n) &\equiv a \tilde A_\mu(x),               \cr
 }}
where $a$ is the lattice spacing, $x_\mu = an_\mu,$
and a tilde denotes the
dimensional continuum quantity. In the same manner, the dimensionless
mass(-matrix) is given by
\eqn\Id{
     M = a\tilde m,
}
which plays the role of an infrared regulator: the
infrared divergence is the
only remaining singularity on the lattice. If the action is chosen properly
(this must be checked since we lift the hermiticity in the following) we can
study the continuum behavior of Feynman integrals by taking $M\rightarrow
0$, that is, $a\rightarrow 0$ with $\tilde m$ being fixed.

For later convenience, we now calculate the Fourier transformations of the
propagator and the vertices. To this end, let us write the action \Ia\ as
\eqn\Ie{
     I = - \sum_{m,n} \bar\psi(m) S^{-1}(m,n) \psi(n),
}
with
\eqn\If{
\eqalign{
  S^{-1}(m,n)\equiv
    & \sum_{\mu,k\geq 1} \Bigl[
      \delta_{m+k\mu,n}\Gamma_\mu^{(+;k)}U_\mu(m)\cdots
U_\mu(m+(k-1)\mu)  \cr
    & + \delta_{m,n+k\mu}\Gamma_\mu^{(-;k)}U_\mu^\dagger(n+(k-1)\mu)
        \cdots U_\mu^\dagger(n) \Bigr]    \cr
    & + \delta_{m,n} \Bigl( \Gamma^{(0)} + M \Bigr).            \cr
}}
We then decompose $S^{-1}(m,n)$ into
\eqn\Ig{
    S^{-1}(m,n) = S_0^{-1}(m,n) + \Sigma(m,n),
 }
where
\eqn\Ih{
    S_0^{-1}(m,n) \equiv S^{-1}(m,n)\vert_{A_\mu=0},
 }
thus
\eqn\Ii{
\eqalign{
   \Sigma(m,n) \equiv
    & \sum_{\mu,k\geq 1} \Bigl\{
      \delta_{m+k\mu,n}\Gamma_\mu^{(+;k)}\bigl[ U_\mu(m)\cdots
U_\mu(m+(k-1)\mu)
                                               - 1 \bigr]        \cr
    & + \delta_{m,n+k\mu}\Gamma_\mu^{(-;k)}\bigl[
U_\mu^\dagger(n+(k-1)\mu)
        \cdots U_\mu^\dagger(n) - 1 \bigr] \Bigr\}.              \cr
   }}
$S_0^{-1}(m,n)$ and $\Sigma(m,n)$ are called the inverse propagator and
 the vertex respectively. Using
\eqn\Ij{
 \psi(n) = \int_p {\rm e}^{ipn}\psi(p),
}
with
 $$
     pn \equiv \sum_\mu p_\mu n_\mu,\quad
     \int_p\equiv \int_{-\pi}^\pi {d^dp \over {(2\pi)^d} },
$$
we get the momentum representation of \Ih;
\eqn\Ik{
  S_0^{-1}(p)=\sum_{\mu,k\geq 1}\Bigl( \Gamma_\mu^{(+;k)}{\rm
e}^{ikp_\mu} +
                  \Gamma_\mu^{(-;k)}{\rm e}^{-ikp_\mu}\Bigr) +  \Gamma^{(0)}
+ M.
     }
Define a divisor ${\cal D}_0 (p)$ such that
\eqn\Ika{
 S_0^{-1}(p) {\cal D}_0 (p) = {\Delta}^{-1} (p) {\bf 1} ,
}
where ${\Delta}^{-1} (p)$ is the scalar inverse propagator. Then
\eqn\Ikb{
S_0(p) = {\Delta} (p){\cal D}_0 (p),
}
which tells us that any singularity of the propagator
$S_0(p)$ is controlled by
${\Delta}(p)$.

The Fourier transformation of the
vertex can be obtained by use of the Taylor expansion with respect to
$A_\mu$ as
 \eqn\Il{
\eqalign{
  & \sum_{m,n}{\rm e}^{-iqm}\Sigma(m,n){\rm e}^{ipn}           \cr
 =& \sum_{N=1}^\infty{1\over N!}\Bigl\{ \sum_{k\geq 1}\Bigl[ \sum_n
    {\rm e}^{i(p-q)(n+k\mu/2)}\bigl(\bar A_\mu^{(k)}(n) \bigr)^N \Bigr]
    v_\mu^{(k;N)}({{p+q}\over 2}) \Bigr \},
  }}
where
 \eqn\Im{
    \bar A_\mu^{(k)}(n) \equiv {1\over k}\sum_{J=0}^{k-1} A_\mu(n+J\mu),
  } and \eqn\In{
    v_\mu^{(k;N)}(p) \equiv {\del^N \over{\del p_\mu^N}}\bigl[
                            \Gamma_\mu^{(+;k)}{\rm e}^{ikp_\mu}
                          + \Gamma_\mu^{(-;k)}{\rm e}^{-ikp_\mu} \bigr].
 }
Inspecting \In\  and \Ik, we can recognize the relationship between the
vertex and the inverse propagator. Furthermore $A_\mu$ is supposed to be a
smooth function under $a\rightarrow 0$;
\eqn\Io{
   A_\mu(n+J\mu) \simeq A_\mu(n) + O(a),  \qquad\quad ( a\rightarrow 0)
}
 to yield instead of \Il,
\eqn\Ip{
\eqalign{
  & \sum_{m,n}{\rm e}^{-iqm}\Sigma(m,n){\rm e}^{ipn}           \cr
 =& \sum_{N=1}^\infty{1\over N!} \sum_n \bigl( A_\mu(n) \bigr)^N
    {\rm e}^{i(p-q)n} v_\mu^{(N)}\Bigl({{p+q}\over 2} \Bigr) + O(a),   \cr
 }}
where
\eqn\Iq{
    v_\mu^{(N)}(p) \equiv {\del^N \over{\del p_\mu^N}} S_0^{-1}(p),
   }
since the $k$'s sum in \Il\ and \Im\ can be performed. \Iq\  is
the lattice Ward relation.

Now we study various situations for the general fermion action \Ia:

\item{(I)} Chiral Invariance (when $M=0$):
\eqn\Ir{
 \{ \gamma_5, \Gamma_\mu^{(\pm;k)} \} =\{ \gamma_5, \Gamma^{(0)} \} =
   0 \qquad
     ( {\rm for\  all}\  k\  \  {\rm and\  } \mu).
}
In this case the inverse propagator is given by  in $d=2$
\eqn\Isa{
\eqalign{
         S_0^{-1}(p) &= i\gamma\cdot F(p) + M   , \cr
    {\Delta}^{-1}(p)& = F^2 + M^2   , \cr
}}
and in $d=4$
\eqn\Isb{
\eqalign{
    S_0^{-1}(p) &= i\gamma\cdot F(p) +\gamma\gamma_5\cdot G(p) + M,  \cr
   {\Delta}^{-1}(p)& = {(F^2 - G^2 + M^2)}^2 + 4M^2G^2+4(F\cdot G)^2  , \cr
   }}
 where use has been made of the abbreviation;
\eqn\It{
\eqalign{
 F^2 &\equiv \sum_\mu {F_\mu
}^2, \quad \quad  \quad \quad  \quad \quad   G^2 \equiv \sum_\mu {G_\mu}^2 ,
\cr
\gamma\cdot F(p) &\equiv \sum_\mu\gamma_\mu F_\mu(p), \quad
  \gamma\gamma_5\cdot G(p) \equiv \sum_\mu\gamma_\mu\gamma_5
G_\mu(p).   \cr
}}
\item{(II)} Hermiticity: the reflection positivity tells us that at
least the following conditions\foot{Additional conditions are
also necessary but
the form of them is not so simple; thus we relegate it
to the appendix.} should be
fulfilled simultaneously:
 \eqn\Iu{
 \eqalign{
\gamma_d{\Gamma_i^{(\pm;k)}}^\dagger\gamma_d &=
\Gamma_i^{(\mp;k)} \qquad  {\rm for}\  \   i=1,2,\cdots, d-1,\cr
\gamma_d{\Gamma_d^{(\pm;k)}}^\dagger\gamma_d &=
\Gamma_d^{(\pm;k)},  \quad
\gamma_d{\Gamma^{(0)}}^\dagger\gamma_d =\Gamma^{(0)} . \cr
 }}
as well as  $k{\leq}1.$

We now impose some conditions to our general action.

\item{(a)} Naive Continuum Limit: the inverse propagator \Ik, \Isa, \Isb\
behaves such that
\eqn\Iv{
    S_0^{-1}(p) \mathop{\rightarrow}\limits_{p\rightarrow 0}
    i\gamma\cdot p + M.
}

\item{(b)} Regularization Free: there should be no singularity in $S_0(p)$
 as long as  $M\not=0$. The condition reads from \Ikb\
\eqn\Iwa{
    |{\Delta}^{-1}(p)| \gtnoteq 0 \qquad {\rm for\  } M^2\gtnoteq 0.
}
In a chirally invariant case, this turns out, according
to \Isa\ and \Isb, to be
\eqn\Iwb{
  | F^2(p)+M^2| \gtnoteq 0,   \qquad {\rm for}\  d=2,
}
or
\eqn\Iwc{
 |{(F^2 - G^2 + M^2)}^2+4M^2G^2+4(F\cdot G)^2| \gtnoteq 0,
\qquad {\rm for}\
d=4.
}

\item{(c)} Pole Singularity: when $M=0$ any singularity of the propagator
$S_0(p)$, that is, of $\Delta (p)$, must be a pole.

\item{(d)} Direction Interchange Symmetry(DIS): the form of
$\Gamma_\mu^{(\pm;k)}$is unchanged after an interchange of
$\mu$-direction.

Here we explain these conditions: the condition (a) is a fundamental
requirement for any lattice model. (b) is especially necessary in
nonhermitian cases since $S_0^{-1}(p)$ may contain complex numbers in
general. Unless this is satisfied, we need an additional regularization
even on the lattice.  Furthermore it is
necessary for the condition (c) to be
fulfilled since otherwise there need additional (and maybe very cumbersome)
methods to estimate the lattice Feynman integral.
As for the condition (d); this
is a statement  of the relativistic invariance on the lattice.
Without this we
have less symmetric model to recover the continuum limit much slower.
(Note that this condition is different from the lattice rotation symmetry.
See below.)

Now we check some explicit examples:

\item{(1)} Naive Dirac Case: we take
 $$
   \Gamma_\mu^{(+;1)} = {\gamma_\mu\over 2}, \quad
   \Gamma_\mu^{(-;1)} = -{\gamma_\mu\over 2}, \quad
    {\rm others}=0.
$$
The action is
\eqn\Ixa{
 \eqalign{
  I^{(1)} = &-\sum_{n,\mu}\Biggl\{ {1\over 2}\Bigl[
                \bar\psi(n)\gamma_\mu U_\mu(n)\psi(n+\mu)
              - \bar\psi(n+\mu)\gamma_\mu U_\mu^\dagger(n)\psi(n) \Bigr]
    \Biggr\}    \cr
             & - \sum_n \bar\psi(n) M \psi(n),     \cr
 }}
 which is, from (I) and (II), chirally invariant and hermitian. The inverse
propagator is given by
  \eqn\Ixb{
  \eqalign{
     S_0^{-1}(p) = & i\gamma\cdot\sin p + M, \cr
    {\Delta}^{-1}(p) = & {(\sin p)}^2 + M^2,   \cr
}}
with ${(\sin p)}^2 \equiv \sum_\mu\sin^2p_\mu.$ $ {\Delta}^{-1}(p)$ has
$2^D$ zeros at $p_\mu = 0$ or $\pi$, around which
\eqn\Ixba{
{\Delta}^{-1}(p)  \mathop{\longrightarrow}\limits_{p_\mu \rightarrow 0,\pi}
    p^2 + M^2 .
}
All the conditions (a) $\sim$ (d) are apparently satisfied.

\item{(2)} Wilson Case: we take
\eqn\Ixc{
\eqalign{
  & \Gamma^{(0)} = rd, \qquad
    \Gamma_\mu^{(+;1)} = {\gamma_\mu-r\over 2},       \cr
  & \Gamma_\mu^{(-;1)} = -{\gamma_\mu+r\over 2}, \quad
    {\rm others}=0, \quad {\rm r \in \bf R}.              \cr
   }}
The action is
\eqn\Ixd{
      \eqalign{ I^{(2)}=&-\sum_{n,\mu}\Biggl\{ \half\Bigl[
                \bar\psi(n)\gamma_\mu U_\mu(n)\psi(n+\mu)
        - \bar\psi(n+\mu)\gamma_\mu U_\mu^\dagger(n)\psi(n) \Bigr]     \cr
        &-{r\over 2}\Bigl[ \bar\psi(n) U_\mu(n)\psi(n+\mu)
                       +\bar\psi(n+\mu)U_\mu^\dagger(n)\psi(n)
                        -2 \bar\psi(n)\psi(n) \Bigr] \Biggr\} \cr
            & - \sum_n \bar\psi(n) M \psi(n).     \cr
 }}
This model is not chirally invariant as long as $r \neq 0$. (When
$r\rightarrow 0$, this becomes to the case (1).) If   $0\leq r^2\leq 1$,
according to the reflection positivity\foot{See the appendix again.},
hermiticity is satisfied. The inverse propagator is
\eqn\Ixe{
  \eqalign{
                     S_0^{-1}(p) = & i\gamma\cdot\sin p + rC(p) + M, \cr
            \Delta^{-1}(p) = &{(\sin p)}^2 + \bigl( M+r C(p) \bigr)^2,\cr
                C(p) &\equiv \sum_\mu(1-\cos p_\mu). \cr
}}
Due to the $C(p)$ term, ${\Delta} ^{-1} (p)$ has only one zero at $p^{(0)} =
(0,0,\dots)$ where
\eqn\Ixea{
    {\Delta}^{-1}(p)  \mathop{\longrightarrow}\limits_{p\rightarrow p^{(0)} }
    p^2 + M^2 .
}

\item{(3)} Alonso-Boucaudo-Cortes-Rivas(ABCR) Model\ref\ABCR{J. -L.
Alonso, P. Boucaud, J. -L. Cortes and E. Rivas,
Phys. Lett. {\bf B201}, 340 (1988);
Phys. Rev. {\bf D40}, 4123 (1989).}: we take
\eqn\Iza{
\eqalign{
           &\Gamma_\mu^{(+;1)} = - {1+i\over 2}\gamma_\mu + {i\over 2d}
                               \sum_\nu \gamma_\nu       \cr
            \Gamma_\mu^{(-;1)} & = {1-i\over 2}\gamma_\mu + {i\over 2d}
                     \sum_\nu \gamma_\nu ,  \qquad  {\rm others}=0. \cr
}}
The action is
\eqn\Izb{
\eqalign{
I^{(3)}=&-\sum_{n,\mu}\Biggl\{ \half\Bigl[
                \bar\psi(n) \Bigl\{ (1+i) \gamma_\mu - {i\over d}\sum_\nu
  \gamma_\nu \Bigr\}  U_\mu(n)\psi(n+\mu)  \cr
       & - \bar\psi(n+\mu) \Bigl\{ (1-i)\gamma_\mu  +  {i\over d}\sum_\nu
   \gamma_\nu \Bigr\} U_\mu^\dagger(n)\psi(n)\Bigr]     \cr
        & - \sum_n \bar\psi(n) M \psi(n),     \cr
 }}
which is hermitian as well as chirally invariant. The inverse propagator is
\eqn\Izc{
  \eqalign{
 S_0^{-1}(p) = & i\gamma\cdot F(p)+ M , \quad  \Delta^{-1}(p) = F^2+M^2,\cr
 F_\mu(p) & \equiv \sin p_\mu  + 1 - \cos p_\mu - {2\over d} C(p),   \cr
}}
whose zeros are
\eqn\Izd{
p^{(0)} = (0,0, \dots), \qquad     p^{(1)} = (\pi/2, \pi/2,\dots) .
}
Around these
\eqn\Izd{
    {\Delta}^{-1}(p)  \mathop{\longrightarrow}\limits_{p\rightarrow p^{(0)},
p^{(1)} }
    p^2 + M^2 .
}
Thus this model satisfies all the conditions (a)
$\sim $ (d). As previously stated,
this model does not have a lattice rotation symmetry \ABCR\ but does
the Direction Interchange Symmetry(DIS).

\item{(4)} Nonhermitian Case (i)\foot{The cases (4)
and (5) are not only simple
but also obtainable from the operator formalism by using fermion coherent
states as is the Wilson cases\ref\OKS{Y. Ohnuki and T. Kashiwa, Prog. Theor.
Phys. {\bf 60}, 548 (1978); T. Kashiwa and
H. So, Prog. Theor. Phys. {\bf73}, 762
(1985)}.}:
\eqn\Ixf{
\eqalign{
  & \Gamma^{(0)} = \sum_\mu \gamma_\mu\gamma_5,\qquad
    \Gamma_\mu^{(+;1)} = {\gamma_\mu(1-\gamma_5)\over 2},       \cr
  & \Gamma_\mu^{(-;1)} = -{\gamma_\mu(1+\gamma_5)\over 2}, \qquad
    {\rm others}=0.              \cr
           }}
The action is
\eqn\Ixg{
\eqalign{
  I^{(4)} = -\sum_{n,\mu}\Biggl\{ &\half\Bigl[
                \bar\psi(n)\gamma_\mu U_\mu(n)\psi(n+\mu)
       - \bar\psi(n+\mu)\gamma_\mu U_\mu^\dagger(n)\psi(n) \Bigr]     \cr
            &-\half\Bigl[
    \bar\psi(n)\gamma_\mu\gamma_5U_\mu(n)\psi(n+\mu)
              + \bar\psi(n+\mu)\gamma_\mu\gamma_5U_\mu^\dagger(n)\psi(n)
    \cr
             &-2 \bar\psi(n)\gamma_\mu\gamma_5\psi(n) \Bigr] \Biggr\} r -
     \sum_n
                      \bar\psi(n) M \psi(n),     \cr
           }}
which is chirally invariant but nonhermitian. The inverse propagator is
given by  in $d=2$
\eqn\Ixh{
\eqalign{
  S_0^{-1}(&p) = \sum_{\mu,\nu} i\gamma_\mu\bigl[ \sin p_\mu
              + \epsilon_{\mu\nu}(1-\cos p_\nu) \bigr]+M , \cr
    \Delta^{-1}(&p)  = \sum_\mu\bigl[\sin p_\mu +
\sum_\nu\epsilon_{\mu\nu}(1-\cos p_\nu)\bigr]^2
     + M^2      \cr
  = 2&(1-\cos p_1)(1-\sin p_2)+2(1-\cos p_2)(1+\sin p_1)+M^2  , \cr
}}
whose zeros (when $M=0$) are
\eqn\Ixha{
       p^{(0)}=(0,0), \qquad  p^{(1)}=(-\pi/2,\pi/2).
}
Around these
\eqn\Ixhb{
 {\Delta}^{-1}(p)\mathop{\rightarrow}\limits_{p \rightarrow
p^{(0)}, p^{(1)}  }   p^2 + M^2 .
}
Thus all the conditions are satisfied. In the four dimensional case
\eqn\Ixhc{
\eqalign{
      S_0^{-1}(p)&= \sum_{\mu} \bigl[ i\gamma_\mu\sin p_\mu
              + \gamma_\mu\gamma_5(1-\cos p_\mu) \bigr]+M ,\cr
  \Delta^{-1}(p) &= (F^2-G^2 + M^2)^2+4M^2G^2+4(F\cdot G)^2 ,
\cr
}}
with $F_\mu=\sin p_\mu$ and $G_\mu=1-\cos p_\mu$. When $M=0$, the zeros
should satisfy
\eqn\Ixhd{
\eqalign{
         &\sum_\mu \cos p_\mu(1-\cos p_\mu) =0,  \cr
         &\sum_\mu \sin p_\mu(1-\cos p_\mu) =0.  \cr
}}
Introducing $x$ and $y$ such that
\eqn\Ixhe{
 p=(x,-x,y,-y) {\rm\  or\  }(x,y,-x,-y){\rm\  or\  }(x,y,-y,-x),
  }
we can see the second equation is trivially satisfied and the first one
becomes
\eqn\Ixhf{
  \cos x(1-\cos x) + \cos y(1-\cos y) = 0,
 }
 to give a circle in $(\cos x, \cos y)$-plane. Thus the zeros of
\Ixhc\  are described by the semicircle defined by
the intersection of \Ixhf\
and the domain $\{ |\cos x|\leq 1\} \cap \{ |\cos y|\leq 1\}$. (See
Fig.1.) As can be read from the figure, $p^{(0)} =(0,0,0,0)$ is isolated
but additional zeros are not poles but rather a cut which breaks the
condition (c).

%%%%%%%%%%%%%%%%%%%%%%%%%%%%%%%%%%%%%%
% Input figure 1 ( for Mac )
%%%%%%%%%%%%%%%%%%%%%%%%%%%%%%%%%%%%%%
%\vskip 1.0truecm
%\special{illustration Fig1m}
%\hskip 3cm\vbox{\hbox{\graph} \hbox{Fig.1:   Allowed region of
%(2.49)} \hbox{Point A corresponds to $p^{(0)} = (0,0,0,0)$. } }
%%%%%%%%%%%%%%%%%%%%%%%%%%%%%%%%%%%%%%
% Input Figure 1 ( by Psfig )
%%%%%%%%%%%%%%%%%%%%%%%%%%%%%%%%%%%%%%
\centerline{\epsffile{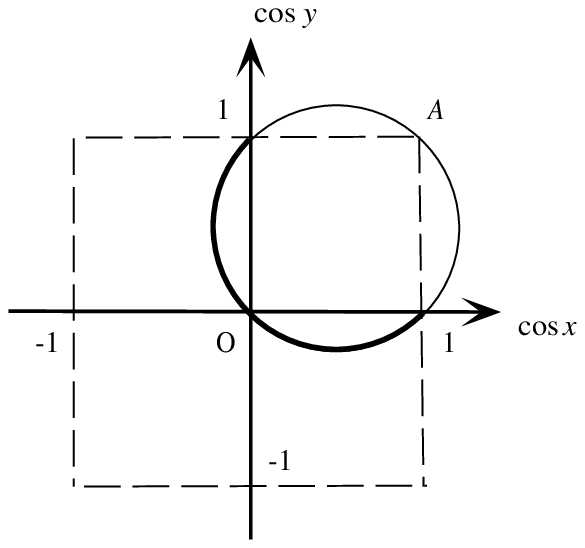}}
\hskip 3cm\vbox{\hbox {Fig.1:   Allowed region of (2.49)}
\hbox{Point A corresponds to $p^{(0)} = (0,0,0,0,)$.}}

\item{(5)} Nonhermitian Case (ii): we consider
$$
   \Gamma^{(0)} = -\sum_\mu \gamma_\mu, \qquad \Gamma_\mu^{(+;1)} =
\gamma_\mu,
    \qquad {\rm others} = 0,
$$
to give
 \eqn\Ixj{
  I^{(5)}=-\sum_{n,\mu} \bar\psi(n)\gamma_\mu\bigl( U_\mu(n)\psi(n+\mu)
             - \psi(n) \bigr) - \sum_n \bar\psi(n) M \psi(n).
      }
This is also chirally invariant and nonhermitian. The inverse propagator
in this case is given by
\eqn\Ixk{
  S_0^{-1}(p)  = \sum_\mu \gamma_\mu\bigl({\rm e}^{ip_\mu} -1 \bigr) + M,
\quad
{\Delta}^{-1}(p)  =   \sum_\mu \bigl( {\rm e}^{ip_\mu} -1 \bigr)^2+M^2 .
}
Then
\eqn\Ixka{
\eqalign{
&|{\Delta}^{-1}(p)|    \cr
          &  =  \sqrt{\Bigl[ \sum_\mu 2\cos p_\mu(1-\cos p_\mu) -
M^2 \Bigr]^2
     + 4\Bigl[ \sum_\mu \sin p_\mu(1-\cos p_\mu)\Bigr]^2},
 }}
which vanishes at $p=(p_0,-p_0)$ with $p_0$ being given by $\cos p_0
=(1 \pm \sqrt{1-M^2})/2$ in $d=2$ for example. Therefore this model has a
singularity even when $M\neq 0$.\par

\newsec{Chiral Ward-Takahashi Identity and Continuum Limit}
In this section, we discuss the chiral Ward-Takahashi
identity and its behavior
in the continuum limit, $a\rightarrow 0$.

The chiral transformation,
\eqn\IIa{
   \psi(n) \mapsto {\rm e}^{i\theta(n)\gamma_5}\psi(n), \qquad
   \bar\psi(n) \mapsto \bar\psi(n){\rm e}^{i\theta(n)\gamma_5},
   } applying to integration variables of the partition function, \eqn\IIb{
  Z[A] \equiv \int\prod_n[d\psi(n)d\bar\psi(n)]
              \exp\bigl[ I[\psi,\bar\psi;A] \bigr],
    }
with $I$ being given by \Ia\  leads us to the Ward-Takahashi identity,
\eqn\IIc{
  \vev{ \Bigl[\bar\psi(n)i\gamma_5{\del\over{\del\bar\psi(n)}}
  + i\gamma_5\psi(n){\del\over{\del\psi(n)}} \Bigr] I } \equiv 0,
 }
where
\eqn\IId{
 \vev{ {\cal O} } \equiv \int\prod_n[d\psi(n)d\bar\psi(n)]{\cal O}{\rm e}^I
                             / Z[A].
 }
Equation \IIc\  can be read, by writing
\eqn\IIe{
 \eqalign{
                 J^{(+)}(n;k)\equiv
          &\sum_\mu \bar\psi(n) i\gamma_5\Gamma_\mu^{(+;k)} U_\mu(n)\cdots
        U_\mu(n+(k-1)\mu)\psi(n+k\mu), \cr
          J^{(-)}(n;k)\equiv
           &\sum_\mu \bar\psi(n+k\mu) i\gamma_5\Gamma_\mu^{(-;k)}
        U_\mu^\dagger(n+(k-1)\mu)\cdots U_\mu^\dagger(n)\psi(n),
         }}
as
 \eqn\IIf{
\eqalign{
    \vev{ \sum_{ k\geq 1 \atop{\rm invariant} } &\Bigl[
          \bigl( J^{(+)}(n;k)-J^{(+)}(n-k;k)\bigr) -
          \bigl( J^{(-)}(n;k)-J^{(-)}(n-k;k)\bigr) \Bigr] }  \cr
    & + \vev{ X(n) }   =  - \vev{ \bar\psi(n) 2iM\gamma_5 \psi(n) },
  }}
where
\eqn\IIg{
\eqalign{
 X(n)\equiv &\sum_{ k\geq 1 \atop{\rm noninvariant} }\Bigl[
           \bigl( J^{(+)}(n;k)+J^{(+)}(n-k;k)\bigr)  \cr
           & + \bigl( J^{(-)}(n;k)+J^{(-)}(n-k;k)\bigr) \Bigr]
      + \bar\psi(n) \{ i\gamma_5, \Gamma^{(0)}\} \psi(n).
\cr
 }      }
Here we have divided the general action into chirally invariant and
noninvariant parts characterized by \Ir\  and
$$
    \bigl[ \gamma_5, \Gamma_\mu^{(\pm;k)}\bigr] = 0, \qquad
    {\rm for\  noninvariant\  pieces}.
$$
Since $k$'s sum is finite, the first term
in the left-hand side of \IIf\  vanishes when being
summed up with respect to
$n$ to yield
\eqn\IIh{
  \sum_n\vev{X(n)}=-\sum_n\vev{\bar\psi(n)2iM\gamma_5\psi(n)}.
}
We require that {\sl this Ward-Takahashi
identity must be fulfilled at every
stage while taking continuum limit, $a\rightarrow 0$} .

In the following, we study the right-hand side of \IIh\  in terms of
an $A_\mu$-expansion. To achieve this, we take the trace with respect to
$\gamma$-matrices as well as the mass matrix
(and also to gauge-group index, if
any) then recall that $S(m,n)$ is given by \Ig\ to find
\eqn\IIj{
\eqalign{
   -\sum_n\vev{\bar\psi(n)2iM\gamma_5\psi(n)}
  =& \sum_n \Tr\bigl[ 2iM\gamma_5 S(n,n) \bigr]     \cr
  =& \sum_n \Bigl[ \Ar^{(0)} +
     \sum_\mu \int_p{\rm e}^{ipn} A_\mu(p)\Ar_\mu^{(1)}(p)       \cr
   & + \half\sum_{\mu,\nu}\int_{p,q} {\rm e}^{i(p+q)n} A_\mu(p) A_\nu(q)
       \Ar_{\mu\nu}^{(2)}(p,q) + O(A^3) \Bigr],                  \cr
  }}
where
\eqn\IIk{
      A_\mu(p) \equiv \sum_n {\rm e}^{-ipn} A_\mu(n), }
and
\eqn\IIl{
\eqalign{
  \Ar^{(0)} \equiv & \int_l \Tr\bigl[ 2iM\gamma_5 S_0(l) \bigr],   \cr
  \Ar_\mu^{(1)}(p) \equiv &
   -\int_l \Tr\Bigl[ 2iM\gamma_5 S_0(l + {p\over 2} )v_\mu^{(1)}(l)S_0(l
-{p\over 2} )
         \Bigr]      ,      \cr
  \Ar_{\mu\nu}^{(2)}(p,q) \equiv &
   -\int_l \Tr\Bigl[ 2iM\gamma_5 S_0(l+{{p+q}\over 2}) \delta_{\mu\nu}
             v_\mu^{(2)}(l) S_0(l-{{p+q}\over 2}) \Bigr]          \cr
  & +\int_l \Tr\Bigl[ 2iM\gamma_5 S_0(l+{{p+q}\over 2}) v_\mu^{(1)}(l+
{q\over 2 }) S_0(l-{{p-q}\over 2})     \cr
  &   \qquad   \qquad \qquad        \times    v_\nu^{(1)}(l-{p\over 2})
                     S_0(l-{{p+q}\over 2}) \Bigr]          \cr
  & + \int_l \Tr\Bigl[ 2iM\gamma_5 S_0(l+{{p+q}\over 2})
v_\nu^{(1)}(l+{p\over 2})
                     S_0(l+{{p-q}\over 2})     \cr
   &   \qquad \qquad \qquad   \times  v_\mu^{(1)}(l-{q\over 2})
                     S_0(l-{{p+q}\over 2}) \Bigr] .          \cr
 }}
We calculate each term of the right-hand side of \IIj\
under $a\rightarrow 0$. To this end we expand each coefficient around $p=0$.
The power counting on the
lattice \ref\KSF{T.  Kashiwa and H. So, Prog. Theor.
Phys. {\bf 73}, 762 (1985); K. Funakubo and
T. Kashiwa, Phys. Rev. {\bf D38},
2602 (1988).} tells that the first
derivative of $\Ar_\mu^{(1)}(p)$ is relevant in
$d=2$ while the second derivative, $\del^2\Ar_{\mu\nu}^{(2)}(p,q) / \del
p_\alpha\del q_\beta$, is in $d=4$.

Let us calculate  $\Ar_\mu^{(1)}(p)$ in the case of Wilson action
\Ixd\ in two dimensions.  Due to the chiral noninvariance, we have
$X(n)$;  \eqn\IIm{
\eqalign{
  X(n) = -{r\over 2}\sum_\mu\Biggl\{
 & \Bigl[ \bar\psi(n)i\gamma_5 U_\mu(n)\psi(n+\mu)
 + \bar\psi(n-\mu)i\gamma_5 U_\mu(n-\mu)\psi(n) \Bigr] \cr
 + \Bigl[ \bar\psi(n+\mu)i & \gamma_5 U_\mu^\dagger(n)\psi(n)
 + \bar\psi(n)i\gamma_5 U_\mu^\dagger(n-\mu)\psi(n-\mu)
        \Bigr] \cr
 & -4 \bar\psi(n)i\gamma_5 \psi(n) \Biggr\}.
   }}
Note that $\Ar^{(0)}=0$ because of the trace property. In view of \Ixe\
we find
\eqn\IIp{
 \Ar_\mu^{(1)}(p) = -2\int_l \Delta(l+{p\over 2})\Delta(l-{p\over 2}) \bigl[
4M^2
         \sum_\nu \epsilon_{\mu\nu}\cos l_\mu \cos l_\nu \sin {p_\nu
\over 2}   + r N_\mu(l;p) \bigr],
 }
where
\eqn\IIq{
 \eqalign{
 N_\mu(l;p) \equiv
     2M&\Biggl\{ \sum_\nu \cos l_\mu \epsilon_{\mu\nu} \Bigl[
           \sin(l+{p\over 2})_\nu C(l-{p\over 2}) - \sin(l-{p\over 2})_\nu
C(l+{p\over 2}) \Bigr]      \cr
     & \qquad\qquad   + \sum_{\nu,\lambda}\sin l_\mu \epsilon_{\nu\lambda}
     \sin(l+{p\over 2})_\nu \sin(l-{p\over 2})_\lambda \Biggr\} .
\cr  }}
We expand $\Ar_\mu^{(1)}(p)$ around $p=0$;
\eqn\IIex{
    \Ar_\mu^{(1)}(p) =   {{\partial {\Ar_\mu}^{(1)}}\over
    {\partial p_\nu \quad} }\Bigg\vert_{p=0} p_\nu
    + O(p^3),
}
where
\eqn\IIr{
\eqalign{
   {{\partial {\Ar_\mu}^{(1)}}\over
   {\partial p_\nu \quad} }\Bigg\vert_{p=0} =
              &-4M^2 \int_l\Delta(l)^2
           \bigl[ \epsilon_{\mu \nu}\cos{l_\mu}\cos{l_\nu }\bigr] \cr
 &-4Mr \int_l\Delta(l)^2 \bigl[ \epsilon_{\mu\nu} \cos{l_\mu} \cos{l_\nu}
                                   C(l)  \cr
    & - \sum_\lambda \bigl ( \epsilon_{\mu\lambda} \sin{l_\nu} \cos{l_\mu}
    - \epsilon_{\nu\lambda} \sin{l_\mu} \cos{l_\nu} \bigr)
                                  \sin{l_\lambda} \bigr] . \cr
} }
In order to estimate the above integrals, we first
recall that there is only one
zero \Ixea\ then divide the integration region into $\De{l^{(0)}}$ where
\eqn\IIo{
        \De{l^{(0)}} \equiv \{ l | (l-l^{(0)})^2 \leq \epsilon^2 \} ,
}
with $l^{(0)} = (0,0)$ and the rest,
$\Dr{l^{(0)}}\equiv [-\pi,\pi]^2-\De{l^{(0)}}$.
We find    $$
 \lim_{M\rightarrow 0} M \int_{\Dr{l^{(0)}}} \bigl [ \cdots \cdots \bigr ]
      = 0,
$$
since there is no singularity under the integration. Also
 \eqn\IIsb{
   \lim_{M\rightarrow 0} M^2 \int_{\De{l^{(0)}}} { {d^2l}\over (2\pi)^2}
              { 1\over(l^2+M^2)^2} = {1\over4\pi},
}
\eqn\IIsa{
   \lim_{M\rightarrow 0} M \int_{\De{l^{(0)}}} { {d^2l}\over (2\pi)^2}
               {l^{2m}\over(l^2+M^2)^2} = 0,  \qquad {\rm for\  } m\geq 1.
}
Using these, we obtain
\eqn\IIsd{
 \eqalign{
   \lim_{M\rightarrow 0} {{\partial{\Ar_\mu}^{(1)}}\over
                   {\partial p_\nu \quad} }\Bigg\vert_{p=0}
    = &\lim_{M\rightarrow 0} \biggl[  -4M^2 \int_{\De{l^{(0)}}} { {d^2l}\over
    (2\pi)^2}{\epsilon_{\mu \nu} \over(l^2+M^2)^2}  \cr
     &-4Mr\int_{\De{l^{(0)}}} { {d^2l}\over (2\pi)^2}
              {\epsilon_{\mu \nu} l^2 -\sum_\lambda \bigl(\epsilon_{\mu
         \lambda} l_\nu - \epsilon_{\nu \lambda}  l_\mu \bigr)
             l_\lambda  \over(l^2+M^2)^2} \biggr] \cr
    = & - {\epsilon_{\mu \nu}\over \pi}.
}}
{}From \IIsd, the right hand side of
\IIh\ becomes, in view of \IIj\ and \IIex, as
\eqn\IIt{
\eqalign{
   \lim_{a\rightarrow 0} \sum_n \vev{ X(n) }
 & = \lim_{a\rightarrow 0} \sum_n \int_p {\rm e}^{ipn} A_\mu(p)
          \bigl( -{1\over\pi}\epsilon_{\mu\nu}p_\nu \bigr)        \cr
 & = i \int d^2x {1\over\pi}\epsilon_{\mu\nu}\del_\nu \tilde A_\mu(x),   \cr
  }}
which reveals the correct anomaly relation
in the continuum (by inserting \IIt\
into \IIf );
$$
  \del_\mu J_{5\mu}(x) = {i\over\pi}\epsilon_{\mu\nu}\tilde F_{\mu\nu}
                          - 2M J_5 (x),
$$
with
$$
   J_{5\mu}(x) \equiv \tilde{\bar\psi}(x) i\gamma_5\gamma_\mu\tilde\psi(x),
\qquad
   J_5(x) \equiv \tilde{\bar\psi}(x) i\gamma_5\tilde\psi(x).
$$\par

Next let us consider what happens if $r=0$,
that is, in the naive Dirac case:
there is no $X(n)$ in the W-T relation \IIh\  to give
\eqn\IIu{
   \sum_n \vev{\bar\psi(n) 2iM\gamma_5 \psi(n) } = 0.
}
$\Ar^{(0)}=0$ as the above. $\Ar_\mu^{(1)}(p)$ is found, by
putting $r\rightarrow 0$ in \IIp, as
\eqn\IIv{
  \Ar_\mu^{(1)}(p) = -4M^2 \sum_\nu \epsilon_{\mu\nu}p_\nu \int_l
\Delta^2(l)
                       \cos l_\mu \cos l_\nu + O(p^3).
}
Significance in this case is, as can
be seen from \Ixba, that there are four poles
in $\Delta(l)$ when $M\rightarrow 0$: $l^{(1)}=%
(0,\pi)$, $l^{(2)}=(\pi,0)$ and
$l^{(3)}=(\pi,\pi)$ other than  $l^{(0)}=(0,0)$.
Around these poles, the form of
$\Delta(l)$ expressed by \Ixba\ is the same but $\cos l_\mu$ and
$\cos l_\nu$ change their sign to give
\eqn\IIw{
  \Ar_\mu^{(1)}(p) = - {1\over\pi} \epsilon_{\mu\nu}p_\nu(1-1-1+1) = 0.
}
Hence \IIu\  is fulfilled. These additional poles are nothing but species
doublers. {\sl Species doublers control the W-T
identity in the (chirally invariant)
naive Dirac case.}

\newsec{General Case with Chiral Symmetry}

Let us discuss the general case with a chiral symmetry. In two
dimensions, the most general chiral invariant propagator is given by \Isa;
\eqn\IIIa{
  S_0^{-1} = i\gamma\cdot F(p) + M.
}
With the use of this, it is easily to see that
$$\Ar^{(0)} = 0, $$ because of the trace property.
While $\Ar_\mu^{(1)}(p)$ is
given
\eqn\IIIb{
\eqalign{
 \Ar_\mu^{(1)}(p) = &-4M^2\sum_{\alpha,\beta}\epsilon_{\alpha\beta} \int_l
                     \Delta(l+p/2) \Delta(l-p/2)                     \cr
                    & \times{ {\del F_\alpha(l)}\over{\del l_\mu} }
                      \bigl[ F_\beta(l+p/2)-F_\beta(l-p/2) \bigr],    \cr
  }}
where
\eqn\IIIc{
     \Delta(l) \equiv \bigl[ F(l)^2 + M^2 \bigr]^{-1} .
  }
 The Taylor expansion with respect to $p$ leads us to
\eqn\IIId{
     \Ar_\mu^{(1)}(p) = -4M^2 \sum_{\alpha,\beta,\nu} \epsilon_{\alpha\beta}
        p_\nu
        \int_l \Delta^2(l) { {\del F_\alpha(l)}\over{\del l_\mu} }
      { {\del F_\beta(l)}\over{\del l_\nu} } + O(p^3).
 }
The propagator $\Delta(l)$ \IIIc\ has
a pole at $l=l^{(0)}\equiv (0,0)$ when
$M=0$ due to the condition (a) and (c) in
section 2. Thus the contribution from
the domain $\De{l^{(0)}}$ is just the same
as \IIsb\ and the condition (b) and
(c) tells us that there might be another contribution
{}from a pole, say $l^{(i)}$:
\eqn\IIIe{
 \eqalign{
 \Ar_\mu^{(1)}(p)
 = -\lim_{M\rightarrow 0}  & 4M^2 \sum_{\alpha,\beta,\nu}
                     \epsilon_{\alpha\beta}
    p_\nu  \biggl[  \int_{\De{l^{(0)}}} {
                              {\delta_{\alpha\mu}\delta_{\beta\nu}}\over
                                                 (l^2+M^2)^2 }  \cr
 & + \sum_i \int_{\De{l^{(i)}}} \Delta^2(l) { {\del F_\alpha(l)}\over{\del
l_\mu} }
           { {\del F_\beta(l)}\over{\del l_\nu} } \biggr]  +   O(p^3)  \cr
 = - {1\over\pi} \epsilon_{\mu\nu}p_\nu
    - \lim_{M\rightarrow 0}
       & 4M^2 \sum_{\alpha,\beta,\nu, i}  \epsilon_{\alpha\beta} p_\nu
  \int_{\De{l^{(i)}}}   \Delta^2(l) { {\del F_\alpha(l)}\over{\del l_\mu} }
    { {\del F_\beta(l)}\over{\del l_\nu} }
    + O(p^3).      \cr
 }}
The Ward-Takahashi identity in this case is also given by \IIu\. But if
there would be no pole we would obtain
$$
   \sum_n \vev{\bar\psi(n) 2iM\gamma_5 \psi(n) } \not= 0,
$$
instead. Hence $\Delta(l)$ \IIIc\ must
have additional pole(s) to cancel the first
term of \IIIe: there should be ` species doublers' even in nonhermitian
cases\GLR\foot{We use quotation marks since it is not necessary for the
propagator to behave as $p^2+M^2$ around the redundant pole(s).}.

Let us study this situation in the explicit example \Ixg, where
\eqn\IIIf{
   F_\mu(p) = \sin p_\mu + \sum_\nu \epsilon_{\mu\nu}(1-\cos p_\nu).
}
Thus in view of \Ixha\ and \Ixhb, we
have two contributions from $l^{(0)}$ and
$l^{(1)}$ to find that
 \eqn\IIIi{
\eqalign{
 \Ar_\mu^{(1)}(p) &= - 4M^2 \sum_\nu \epsilon_{\mu\nu} p_\nu \int_l
\Delta^2(l)
     \bigl(\cos l_\mu\cos l_\nu + \sin l_\mu\sin l_\nu \bigr)\cr
     &=- 4M^2 \sum_\nu \epsilon_{\mu\nu} p_\nu \Bigl[
     \int_{\De{l^{(0)}}}{ 1\over(l^2+M^2)^2 } +
     \int_{\De{l^{(1)}}}{ -1\over(l^2+M^2)^2 }      \cr
     &\qquad\quad + \int_{D_r} \Delta^2(l)
     \bigl(\cos l_\mu\cos l_\nu + \sin l_\mu\sin l_\nu \bigr)
     \Bigr]     \cr
     &= 0.   \cr
   }}
In this case, we thus find a species doubler.

The situation is the same as in the hermitian case, \Izb, where $F_\mu(p)$
is given by \Izc\ then
\eqn\IIIj{
\eqalign{
 \Ar_\mu^{(1)}(p) = & - 4M^2 \sum_{\alpha,  \beta,  \nu}\epsilon_{\alpha
\beta}
   p_\nu \int_l \Delta^2(l)
   \bigl\{ \delta_{\alpha \mu} (\cos l_\mu  + \sin l_\mu) -{2\over d}
              \sin l_\mu \bigr\}\cr
   &\qquad \qquad \qquad \qquad \times   \bigl\{ \delta_{\beta \nu} (\cos
l_\nu  + \sin l_\nu)
  -{2\over d}   \sin l_\nu \bigr\}   \cr
  =  & - 4M^2 \sum_\nu \epsilon_{\mu\nu} p_\nu
                     \int_{\De{l^{(0)}}}{ 1\over(l^2+M^2)^2 } \cr
   & - 4M^2 \sum_{\alpha, \beta, \nu} \epsilon_{\alpha \beta} p_\nu
\int_{\De{l^{(1)}}}{ \delta_{\alpha \mu} \delta_{\beta \nu} -
\delta_{\alpha \mu} - \delta_{\beta \nu} +1\over(l^2+M^2)^2 }  \cr
=& 0.  \cr
}}
Here the pole at $(\pi/2, \pi/2)$ cancels
the contribution from $p^{(0)}$, as it
should be. This has also a species doubler. But the number
of species doublers is
reduced compare to the naive Dirac case which has $2^2=4$ poles.

\newsec{Discussion}

The discussion in the previous
sections shows that any chirally invariant
model must contain species doubler(s)
provided the theory is well-regurarized
in view of the condition (b). Although our conclusion
has been checked in a
two-dimensional model, it is straightforward to extend
our scenario to four or
higher dimensions.

As far as the number of species doubler(s) is concerned,
the nonhermitian (in $d=2$) and ABCR models(in $d=2, 4 $ and $6 $\ABCR) are
most economical, since they have only one doubler. However if the condition
(d), Direction-Interchange-Symmetry(DIS), is dropped, there open many
possibilities to have less
doubler\ref\K{L. H. Karsten, Phys. Lett. {\bf 104B}, 315
(1981)} \ref\WZ{F. Wilczek, \PRL{59}, 2397 (1987)}.
But if we do expect a better
recovery of the Lorentz covariance together with an
aethetic point of view, we do not adopt the model which breaks the DIS.

There might be many options which has the chiral symmetry, but in order to
study whether the model is workable or not, it should be
carefully checked that
the model is well-regularized (the condition (b) ) together with the
condition
(c); otherwise we may encounter the computational trouble.

In a chirally invariant model, if we give up
the gauge invariance on the lattice,
we have the anomaly.(See \ABCR\ for example.) However we do think that on
the lattice the gauge invariance should be kept all
the time; otherwise we do
lose the guiding principle for building up the lattice model.

\bigbreak\bigskip\centerline{{\bf Acknowledgements}} \nobreak

The authors are grateful to Jan Smit for
informing us an earlier reference of
Karsten, Wolfgang Bock and Maarten Golterman for fruitful
suggestions. T. K.
also thanks Don Petcher, Sergei Zenkin and Istvan Montvay for discussions.

\appendix{A}{On Reflection Positivity}

In this appendix, we summarize the definition and consequences of
reflection positivity and derive restrictions on the general class
of free fermion actions \Ia.

We call the $d$-th direction of the euclidean spacetime the `time'
direction and suppose the time coordinate takes half-odd integer. Let
$\Aa^+$ be the set of funcitons at positive times which take their
values in the Grassmann algebra generated by $\{ \psi(n), \bpsi(n) \; |\;
n_d > 0 \}$.\foot{Although we consider free fermions, the inclusion of the
link variables is straightforward.} The
set $\Aa^-$ is similarly defined with
$n_d<0$.  Introduce an antilinear map $\Theta : \Aa^+ \rightarrow \Aa^-$
defined as\ref\OS{K. Osterwalder and E. Seiler, \AP{110} (1978) 440.}
\eqn\Va{
\eqalign{ & \Theta f(n) = [f(\vartheta n)]^*,   \cr
          & \Theta \psi(n) = \bpsi( \vartheta n ) \gamma_d,   \cr
          & \Theta \bpsi(n) = \gamma_d \psi( \vartheta n ),   \cr
          & \Theta[AB] = (\Theta B)(\Theta A),  \cr
        }
}
where $\vartheta n \equiv ( n_1,n_2,\ldots,-n_d )$ and the asterisk
denotes the complex conjugate. Especially for a fermion bilinear form,
\eqn\Vb{
   \Theta\bigl[ f(l) \bpsi_\alpha(m) \chi_\beta(n) \bigr]
 = f(\vartheta l)^*
         \bigl(\bar\chi(\vartheta n) \gamma_d \bigr)_\beta
         \bigl(\gamma_d \psi(\vartheta m) \bigr)_\alpha ,
}
where $\alpha$ and $\beta$ stand for the spinor and flavor indices.\par
Next we define a set $\cal P$ as the convex cone generated by
$\bigl\{ (\Theta A) A \; | \; A \in \Aa^+ \bigr\}$.
If the action $I$ satisfies $\exp(I) \in \cal P$, the physical
Hilbert space with positive-definite metric can be induced by use
of the functional integral on the lattice\OS. Here we call an action, $I$,
to be refletion positive, when $\exp(I) \in \cal P$.
We denote the entire cubic lattice as $\Lambda$ and divide it into the
positive-time lattice, $\Lambda_+$, and the negative-time one, $\Lambda_-$.
Any action which satisfies locality can always be decomposed as
\eqn\Vc{
     -I = I^{(+)} + I^{(-)} + \Delta I,
}
where $I^{(\pm)}$ belongs to $\Aa^\pm$ and $\Delta I$ is a sum of
products of the fields on $\Lambda_+$ and those on $\Lambda_-$.
Note that $\cal P$ is a multiplicative cone,
{\it i.e.}, if $A$ and $B$ are in
$\cal P$, then $AB \in \cal P$. Hence, sufficient condition for $I$ to be
reflection positive is
\item{(a)} $I^{(-)} = \Theta I^{(+)}$,
\item{(b)} $\Delta I \in \cal P$,\par\noindent
since (a) means ${\rm e}^{ I^{(+)}+I^{(-)} } = \bigl(\Theta
{\rm e}^{I^{(+)}}\bigr){\rm e}^{I^{(+)}} \in \cal P$ and (b) guarantees
the rest; ${\rm e}^{\Delta I}\in \cal P$.\par
Now we derive restrictions on the general class of fermion actions \Ia\
{}from the reflection positivity.
The mass term,  $\sum_n \bpsi(n) M \psi(n)$, is
in $\cal P$ as far as $M$ is hermitian,
so that we put $M=0$ for simplicity. Any
action in this class can be decomposed into two terms,
one of which contains
timelike differences and the rest;  \eqn\Vd{
      -I = I_t + I_s,
}
and further each part can be divided into pieces like \Vc;
\eqn\Ve{\eqalign{
 I_t &= I_t^{(+)} + I_t^{(-)} + \Delta I,    \cr
 I_s &= I_s^{(+)} + I_s^{(-)}.               \cr
     }
}
where
\eqn\Vf{
\eqalign{
 I_t^{(+)}&=   \sum_{k\geq 1}\sum_{\{n_d\geq 1/2 \}}\Bigl[
               \bpsi(n) \Gamma^{(+;k)}_d \psi(n+k\cdot d) +
               \bpsi(n+k\cdot d) \Gamma^{(-;k)}_d \psi(n) \Bigr],   \cr
 I_t^{(-)}&=   \sum_{k\geq 1}\sum_{\{ n_d\leq -k-1/2 \}}\Bigl[
               \bpsi(n) \Gamma^{(+;k)}_d \psi(n+k\cdot d) +
               \bpsi(n+k\cdot d) \Gamma^{(-;k)}_d \psi(n) \Bigr],   \cr
 \Delta I &=   \sum_{k\geq 1}\sum_{\{ -k+1/2\leq n_d\leq -1/2 \}}\Bigl[
               \bpsi(n) \Gamma^{(+;k)}_d \psi(n+k\cdot d) \cr
 &\qquad \qquad \quad \qquad + \bpsi(n+k\cdot d) \Gamma^{(-;k)}_d \psi(n)
\Bigr],   \cr
         }
}
and
\eqn\Vg{
\eqalign{
 I_s^{(+)}&=   \sum_{n_d\geq 1/2} \biggl[ \sum_{i=1}^{d-1} \sum_{k\geq
1}\Bigl[
              \bpsi(n) \Gamma^{(+;k)}_i \psi(n+k\cdot i) +
              \bpsi(n+k\cdot i) \Gamma^{(-;k)}_i \psi(n) \Bigr]   \cr
& \quad \quad \quad \quad + \bpsi(n) \Gamma^{(0)}\psi(n) \biggr] , \cr
I_s^{(-)}&=  \sum_{n_d\leq -1/2} \biggl[ \sum_{i=1}^{d-1} \sum_{k\geq
1}\Bigl[
              \bpsi(n) \Gamma^{(+;k)}_i \psi(n+k\cdot i) +
              \bpsi(n+k\cdot i) \Gamma^{(-;k)}_i \psi(n) \Bigr]   \cr
& \quad \quad \quad \quad + \bpsi(n) \Gamma^{(0)}\psi(n) \biggr] , \cr
        }
}

For $I$ to be reflection positive,
the following conditions must be fulfilled:
\item{(i)}    $I_s^{(-)} = \Theta I_s^{(+)}$,
\item{(ii)}   $I_t^{(-)} = \Theta I_t^{(+)}$,
\item{(iii)}  $\Delta I$ is expressed as
$\Delta I = \sum_j (\Theta A_j)A_j$
              with some $A_j \in \Aa^+$.\par\noindent
According to the definition \Va, the condition (i) is satisfied iff
\eqn\Vh{
 \gamma_d {\Gamma_i^{(\pm;k)}}^\dagger \gamma_d =
\Gamma_i^{(\mp;k)}(i = 1, \ldots, d-1;\  k\geq 1) ; \quad \gamma_d
{\Gamma^{(0)}}^\dagger \gamma_d = \Gamma^{(0)},
}
and the condition (ii) is iff
\eqn\Vi{
 \gamma_d {\Gamma_d^{(\pm;k)}}^\dagger \gamma_d =
\Gamma_d^{(\pm;k)}.
  \qquad\qquad ( k\geq 1)
}
The consequence of the condition (iii) applied to the general class of
actions is that any difference should be less than one. This can be shown
as follows: each term in the right-hand side
of $\Delta I$ in \Vf\ is composed of
$\psi$ on $\Lambda_+$(or $\Lambda_-$) and $\bpsi$ on $\Lambda_-$(or
$\Lambda_+$). Except for the $k=1$ term, any term is `asymmetric' with
respect to the reflection, {\it i.e.},
cannot be written in the form $(\Theta
A)A$ with some $A$ on either $\Lambda_+$ or $\Lambda_-$. There might
remain a possibility that sum of the terms, not one of them, could be
written as  $\sum_j (\Theta A_j)A_j$ where $A_j$ is a linear combination
of $\psi$ on $\Lambda_+$ or that of $\bpsi$ on $\Lambda_+$.

Let $K$ be the maximum integer such that any $\Gamma_d^{(\pm;k)}$ with
$k>K$ vanishes. If $\Delta I$ meets the condition
(iii), it is generally expressed
by introducing a set of elements in $\Aa^+$;
\eqn\Vk{
\eqalign{
 &A_i = \sum_{n_d=1/2}^{K-1/2} \alpha_i(n_d)\psi(n_d),            \cr
 &B_j = \sum_{n_d=1/2}^{K-1/2} \bpsi(n_d)\beta_j^\dagger(n_d),    \cr
        }
}
as
\eqn\Vj{
   \Delta I = \sum_{i=1}^I (\Theta A_i)A_i + \sum_{j=1}^J (\Theta B_j)B_j,
}
where $I$ and $J$ are some finite numbers and $\alpha_i$ and $\beta_j$ are
complex matrices in spinor (and flavor) space. Here we have suppressed the
spatial coordinates. From \Vj, we have
\eqn\Vl{
\eqalign{
 &\sum_i \gamma_d \alpha_i^\dagger(-n_d)\alpha_i(n_d+k) =
\Gamma_d^{(+;k)}, \cr
 &{\rm for}\  -k+1/2\leq n_d\leq -1/2\  \quad  {\rm with}\  1\leq k\leq K,
\cr
 &\sum_i \gamma_d \alpha_i^\dagger(-n_d)\alpha_i(n_d+k+K) = 0,       \cr
 &{\rm for}\  -K+1/2\leq n_d\leq -k-1/2\  \quad {\rm with}\  1\leq k\leq K-1
\  (K\geq 2),    \cr
 }
}
and similar equations for $\beta_j$ and $\Gamma_d^{(-;k)}$. The second
equation requires all $\alpha_i$ but $\alpha_i(1/2)$ vanish. Hence the
difference in the time direction should be less than one;
$\Gamma_d^{(\pm;k)} =0; k \geq 2 $. Further the remaining
$\Gamma_d^{(\pm;1)}$ must be written as
\eqn\Vm{
\eqalign{
 &\sum_i \alpha_i^\dagger\alpha_i = \gamma_d \Gamma_d^{(+;1)},   \cr
 &\sum_j \beta_j^\dagger\beta_j = - \Gamma_d^{(-;1)} \gamma_d,    \cr
         }
}
with some complex matrices $\alpha_i$ and $\beta_j$.
This condition, applied to the Wilson fermion action, requires that
$(1\pm r\gamma_d)/2$ must be positive semi-definite.
It immediately means that $ | r | \leq 1 $, since the eigenvalues of
$(1\pm r\gamma_d)/ 2$ are either $(1+r)/ 2$ or $(1-r)/ 2$.

\listrefs

\bye